\documentclass{svmult}

\usepackage[ukrainian,american]{babel} 
\usepackage[T2A,T1]{CJKutf8}

\usepackage{amssymb, bm}
\usepackage{color}

\usepackage{makeidx}         
\usepackage{graphicx}        
\usepackage{multicol}        
\def\0{\phantom{0}}

\newcommand{\ukr}[1]{\bgroup\fontencoding{T2A}\foreignlanguage{ukrainian}{#1}\egroup}

\begin{document}

\title*{Development of models and methods for the molecular simulation of large systems and molecules}
\titlerunning{Development of models and methods for large systems}
\author{Jonathan Walter\inst{1}\and Thorsten Merker\inst{1}\and Martin Horsch\inst{2}\and Jadran Vrabec\inst{2}\and Hans Hasse\inst{1}}
\authorrunning{Walter, Merker, Horsch, Vrabec, Hasse}
\institute{Lehrstuhl f\"ur Thermodynamik, Technische Universit\"at Kaiserslautern, Erwin-Schr\"odinger-Str. 44, 67663 Kaiserslautern, Germany
\texttt{hans.hasse@mv.uni-kl.de}
\and Institut f\"ur Verfahrenstechnik, MB/ThEt, Universit\"at Paderborn, Warburger Str.\ 100, 33098 Paderborn, Germany}
%
%
\maketitle

\section{Introduction}

The most important factor for quantitative results in molecular dynamics simulation are well developed force fields and models. In the present work, the development of new models and the usage of force fields from the literature in large systems are presented. Both tasks lead to time consuming simulations that require massively parallel high performance computers. In the present work, new models for carbon dioxide \cite{Merker_2010_1} and cyclohexanol \cite{Merker_2010_2} were developed and a new method for the model development is introduced. Force fields and models for the simulation of PNIPAAm hydrogel in pure water and sodium chloride solution were tested and verified \cite{Walter_2010_2} and used in simulations of nucleation processes \cite{Horsch_1,Horsch_2}.\\
The simulations for all investigations were performed on the high performance computer HP XC 4000 at the Steinbuch Centre for Computing in Karlsruhe (Germany), which is equipped with Opteron 2.6 GHz Dual Core processors.

\section{Development of models}
\label{dev}

Molecular models for applications in engineering are parameterized based on quantum mechanical (QM) ab initio calculations and
thermodynamic data. Both models presented here are rigid. For their usage in the engineering field, it is important to be able to describe all thermodynamic properties of the substance quantitatively. As the development of models is rather time consuming, a new and fast procedure is introduced.

\subsection{Carbon dioxide}

\begin{figure}[h!]
  \centering
  \includegraphics[width=0.8\textwidth]{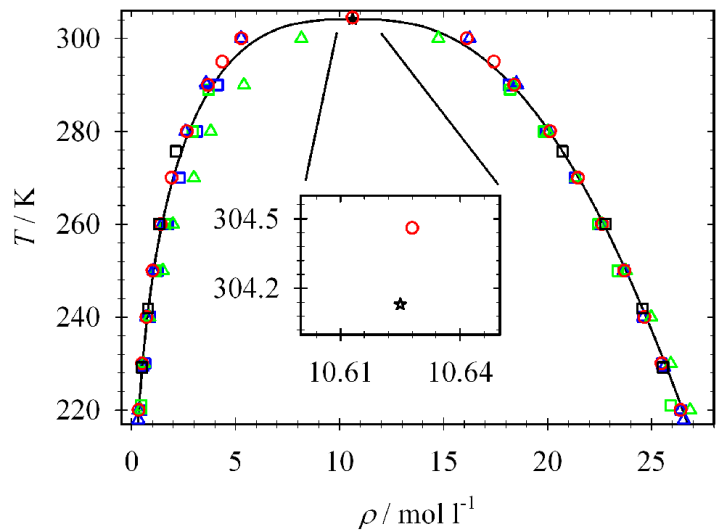}
  \caption{Saturated densities. Simulation results:\textcolor{red}{\large $\circ$}~this work, \textcolor{green}{\small $\square$}~EPM2 \cite{Harris1995}, \textcolor{blue}{\small $\vartriangle$}~Vrabec et al. \cite{Vrabec2001}, \textcolor{blue}{\small $\square$}~Zhang and Duan \cite{Zhang2005,Merker2008}, \textcolor{green}{\small $\vartriangle$}~BBV \cite{Bock2000,Bratschi2007}, {\small $\square$}~M\"{o}ller and Fischer \cite{fischer1994}, {---}~EOS \cite{wagner}, {\small $\bigstar$}~experimental critical point \cite{wagner}. The inset is a magnified view of the critical point.}
  \label{carbon}
\end{figure}

A molecular model for carbon dioxide was developed by optimizing the parameters of the Lennard-Jones sites, the bond length and the quadrupole moment to experimental vapor-liquid equilibrium data. The resulting molecular model are listed in Table \ref{table1}. After adjustment to these properties, it shows mean unsigned deviations to the experiment over the whole temperature range from triple point to critical point of 0.4~\% in saturated liquid density, 1.8~\% in vapor pressure, and 8.1~\% in enthalpy of vaporization. The molecular model is assessed by comparing predicted thermophysical properties with experimental data and a reference equation of state for a large part of the fluid region. The average deviation for density and residual enthalpy is 4.5 and 1.7~\%, respectively. The model is also capable to predict the radial distribution function, the second virial coefficient and transport properties, where the average deviations are 12~\%.
\begin{table}[t!]
\centering
\caption{Parameters of the rigid carbon dioxide model.}
\label{table1}
\begin{tabular}{cccccc}
\hline\noalign{\smallskip}
$\sigma_\mathrm{C}$ & $\epsilon_\mathrm{C}$ & $\sigma_\mathrm{O}$ & $\epsilon_\mathrm{O}$ & $Q$ & $r_\mathrm{C-O}$ \\
nm & kJ$\cdot$mol$^{-1}$ & nm & kJ$\cdot$mol$^{-1}$ & C$\cdot$nm$^{2}$ & nm \\
\noalign{\smallskip}\hline\hline\noalign{\smallskip}
0.28137 & 0.10287 & 0.29755 & 0.83555 & 1.35891$\cdot10^{-21}$ & 0.12869 \\
\noalign{\smallskip}\hline
\end{tabular}
\end{table}\\
As an example, the description of the saturated densities by the model is plotted in Figure \ref{carbon} in comparison to other models and experimental data. Here, the quantitatively good results can be seen.
\\
A more detailed discussion of the development and the results is given in Merker et al. \cite{Merker_2010_1}.

\subsection{Cyclohexanol}

\begin{figure}[h!]
  \centering
  \includegraphics[width=0.8\textwidth]{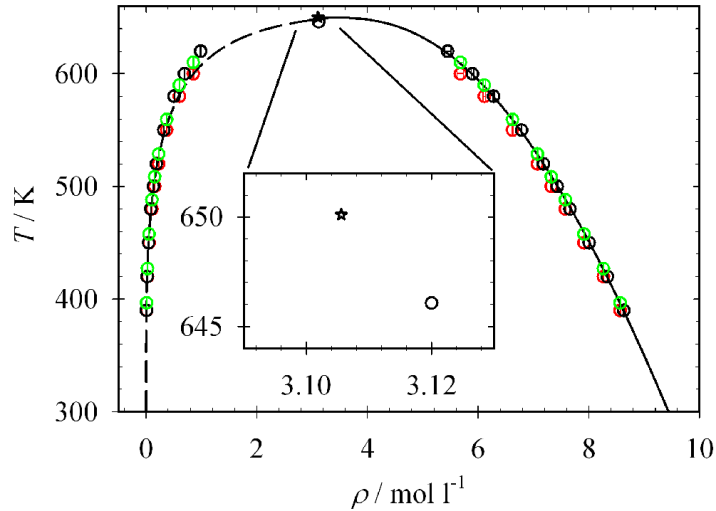}
  \caption{Saturated densities of cyclohexanol: \textcolor{red}{\large $\circ$}~model after first step of optimization, \textcolor{green}{\large$\circ$}~model after reduced unit method, {\large$\circ$}~final model, {---}~DIPPR correlation \cite{DIPPR2005}, $\star$~experimental critical point \cite{DIPPR2005}. Inset: Magnified view on the critical point.}
  \label{cyclo}
\end{figure}

For the development of the cyclohexanol model, a new procedure for parameter adjustment to thermodynamic data via reduced units is introduced. The resulting parameters can be seen in Table \ref{table2}. Compared to experimental data, this cyclohexanol model shows mean unsigned errors of 0.2~\% in saturated liquid density and 3~\% in vapor pressure over the whole temperature range from triple point to critical point. The model was used to predict the second virial coefficient and the transport properties, the average deviations from experimental data are 0.1~l/mol and 25~\%, respectively.
\begin{table}[b!]
\centering
\caption{Coordinates and parameters of the LJ sites and the point charges in the principal axes system of the new molecular model for cyclohexanol. Bold characters indicate the represented atoms.}
\label{table2}
\begin{tabular}{lr@{.}lr@{.}lr@{.}lccr@{.}l} 
\hline\noalign{\smallskip}
Interaction & \multicolumn{2}{c}{$x$} & \multicolumn{2}{c}{$y$} & \multicolumn{2}{c}{$z$} & $\sigma$ & $\epsilon$ & \multicolumn{2}{c}{$q$} \\
Site & \multicolumn{2}{c}{nm} & \multicolumn{2}{c}{nm} & \multicolumn{2}{c}{nm} & nm & kJ$\cdot$mol$^{-1}$ & \multicolumn{2}{c}{$e$} \\ 
\noalign{\smallskip}\hline\hline\noalign{\smallskip}
\textbf{CH}$_{2}$(1) & -0&22118 & -0&05446 & \multicolumn{2}{c}{0} & 0.3412 & 0.86496 & \multicolumn{2}{c}{-} \\
\textbf{CH}$_{2}$(2) & -0&13580 & 0&04614 & -0&15759 & 0.3412 & 0.86496 & \multicolumn{2}{c}{-} \\
\textbf{CH}$_{2}$(3) & 0&05253 & -0&03785 & -0&15617 & 0.3412 & 0.86496 & \multicolumn{2}{c}{-} \\
\textbf{CH}$_{2}$(4) & 0&05253 & -0&03785 & 0&15617 & 0.3412 & 0.86496 & \multicolumn{2}{c}{-} \\ 
\textbf{CH}$_{2}$(5) & -0&10196 & 0&03619 & 0&15759 & 0.3412 & 0.86496 & \multicolumn{2}{c}{-} \\ 
\textbf{CH} & 0&10680 & 0&03619 & \multicolumn{2}{c}{0} & 0.3234 & 0.50727 & 0&27802 \\ 
\textbf{OH} & 0&24134 & 0&00354 & \multicolumn{2}{c}{0} & 0.3150 & 0.71920 & -0&64417 \\ 
\textbf{H}-O & 0&24690 & -0&09368 & \multicolumn{2}{c}{0} & - & - & 0&36615 \\
\noalign{\smallskip}\hline
\end{tabular}
\end{table}\\
As an example, the description of the saturated densities by the model is plotted in Figure \ref{cyclo} in comparison to experimental data. Here, the quantitatively good results can be seen.
\\
For further insight in the development and the new procedure for model development as well as the results see Merker et al. \cite{Merker_2010_2}.

\section{Verification of models}
\label{ver}

For large molecules with internal degrees of freedom, the models get more complex. Here, the usability of different force fields for the description of the swelling of Poly(N-isopropylacrylamide) (PNIPAAm) hydrogel was studied in water as well as in sodium chloride solutions. Different force fields for PNIPAAm and models for sodium chloride from the literature were tested and verified.\\
The other field of investigation where models of different fluids were used, is the simulation of nucleation processes. Here, large systems with a large number of molecules have to be simulated. In order simulated the nucleation rate, good molecular models for the used fluids are important as well.

\subsection{Swelling of poly(N-isopropylacrylamide) hydrogels}
\label{hydrogel}

Hydrogels are three-dimensional hydrophilic polymer networks. Their most characteristic property is their swelling in aqueous solutions by absorbing the solvent, which is influenced by various factors. Hydrogels can be used in many applications. Superabsorbers such as in diapers and contact lenses are the most common applications of hydrogels.\\
The hydrogel which is studied in the present work is built up of poly(N-isopropylacrylamide) (PNIPAAm) cross-linked with N,N'-methylenebisacryl-amide (MBA). The degree of swelling in equilibrium of PNIPAAm is significantly influenced by many factors \cite{Walter_2010_2,Huether_2002,Huether_2006}. On the one hand, the swelling depends on the structure of the hydrogel itself, like the type of the monomer, but also the amount and type of cross-linker and of co-monomers. On the other hand, the environment conditions like temperature, type of solvent, solvent composition, salt concentration or pH-value of the solvent influence the swelling behavior. Varying these factors, the hydrogel typically shows a region where the hydrogel is swollen and a region where it is collapsed. In between those two regions lies the region of conformation transition. The solvent composition or temperature, which is characteristic for that transition, is labeled here with $\Theta$ ($\Theta$-solvent and $\Theta$-temperature). The $\Theta$-conditions are typically only weakly dependent on the amount of cross-linker. Therefore, the $\Theta$-conditions mainly depend on the environmental factors and the nature of the polymer chain.\\
With molecular simulation it should in principle be possible to predict the swelling of different hydrogels upon varying any environmental factor. Two levels of model detail are applied for molecular simulations of hydrogels: coarse-grained and atomistic models. The advantage of coarse-grained models, especially in combination with implicit solvents, is their comparatively low computational cost. Coarse-grained models were used to study conformations of model hydrogels, depending on the solvent or the amount of ions in the solvent and the charges on the hydrogel \cite{Mann_2006,Limbach_2003}. In these studies, it was observed that the hydrogel and the corresponding single polymer chain have similar conformations in good, bad and $\Theta$-solvents: Independent on whether a polymer chain is part of a hydrogel or a single chain, it is collapsed in bad solvents and stretched in good solvents and the transition occurs roughly at the same conditions.\\
Atomistic simulations are computationally much more expensive than coarse-grained simulations, especially when the solvent is modeled explicitly. With atomistic simulations, the dynamic properties and the conformation of the solvent and the monomers of the hydrogels have been studied by several authors \cite{Chiessi_2007,Toensing_2001,Longhi_2004}.\\
In the literature, predominantly model hydrogels were studied. Even for simulations of real hydrogels, the validation on the basis of experimental data was difficult or not attempted at all.\\
In this work, the swelling of PNIPAAm hydrogel was inestigated by atomistic molecular dynamics simulation and by experiment. The results from simulation and experiment are compared to study whether the swelling behavior of hydrogels can be quantitatively predicted by molecular simulation.\\
From experimental data and coarse-grained simulation, it is known that the amount of cross-linker typically has no significant influence on the $\Theta$-conditions of the hydrogel \cite{Huether_2006,Mann_2006,Limbach_2003,Caykara_2006}. The $\Theta$-transition can be understood by studying single PNIPAAm chains as it depends on the interactions between the single chain and the solvent \cite{Walter_2010_2}. In the present work, therefore molecular simulations were performed for single chains in explicitly modeled solvent molecules for a real hydrogel. Namely, the temperature and sodium chloride concentration dependence of the swelling behavior of PNIPAAm in pure water was studied.\\
For the present study, the force fields from the literature were used as published. No parameters were fitted. The results were compared to experimental data on the degree of swelling of PNIPAAm hydrogel in pure water as a function of temperature and concentration of sodium chloride.\\

\subsubsection{Methods and molecular models}

For the molecular dynamics simulations of PNIPAAm in water, the following force fields from the literature were used to describe PNIPAAm: GROMOS-87 (G87) \cite{Gunsteren_1987}, GROMOS-96 53A6 (G53A6) \cite{Oostenbrink_2004_1} and OPLS-AA (OPLS) \cite{Jorgensen_1996}. They were combined with the water model SPC/E \cite{Berendsen_1987}. For the simulations with salt, three different sodium chloride models were used: Chandrasekhar et al. \cite{Chandrasekhar_1984}, OPLS which uses the natrium ion model of Aqvist \cite{Aqvist_1990} and the chloride ion model of Chandrasekhar et al. \cite{Chandrasekhar_1984} as well as G53A6 \cite{Oostenbrink_2004_1}.\\
Here, only the results for the force field combination OPLS + SPC/E are shown, which describes the experimental data best. For results for the other force fields, cf. Walter et al. \cite{Walter_2010_2}. The Lennard-Jones and point charge parameters of the PNIPAAm force field are listed in Table \ref{table4}. The Lennard-Jones sites are characterized by the size parameter $\sigma$ and the energy parameter $\epsilon$.
\begin{table}[t!]
  \centering
  \caption{Lennard-Jones parameters ($\sigma$ and $\epsilon$) and point charge magnitude ($q_{el}$) of the PNIPAAm force field OPLS \cite{Jorgensen_1996}, where e is the elementary charge.}
  \label{table4}
  \begin{tabular}{cccr@{.}l}
    \hline\noalign{\smallskip}
    site & \multicolumn{4}{c}{OPLS} \\
    \noalign{\smallskip}\hline \hline\noalign{\smallskip}
    & $\sigma$ / nm & $\epsilon$ / kJ$\cdot$mol$^{-1}$ & \multicolumn{2}{c}{$q_{el}$ / e} \\
    \noalign{\smallskip}\hline\noalign{\smallskip}
    C & 0.375 & 0.4393 & 0&50 \\
    O & 0.296 & 0.8786 & -0&50 \\
    N & 0.325 & 0.7113 & -0&50 \\
    H & - & - & 0&30 \\
    CH(-N) & 0.350 & 0.2761 & 0&14 \\
    CH & 0.350 & 0.2761 & -0&06 \\
    CH$_{2}$ & 0.350 & 0.2761 & -0&12 \\
    CH$_{3}$ & 0.350 & 0.2761 & -0&18 \\
    H in CH$_{x}$ & - & - & 0&06 \\
    \noalign{\smallskip}\hline
  \end{tabular}
\end{table}\\
For the unlike Lennard-Jones pair interaction, a geometric mean combining rule for both $\sigma$ and $\epsilon$ was used
\begin{equation}
  \sigma_{ij}=\sqrt{\sigma_{i}\cdot\sigma_{j}}, \\
\end{equation}
\begin{equation}
  \epsilon_{ij}=\sqrt{\epsilon_{i}\cdot\epsilon_{j}}.
\end{equation}
For the intramolecular interactions, the method of the 1-4 interactions \cite{Spoel_2005_2} was employed.\\ 
Molecular simulations of PNIPAAm single chains were carried out with versions 4.0.x of the GROMACS simulation package \cite{Hess_2008}. The GROMACS code is optimized for single processors and also for massively parallel machines. It was developed for the simulation of large molecules in solutions.\\
Simulations at temperatures between 280 and 360~K were performed, namely at 290, 300, 310, 320 and 340~K. These allow to obtain the $\Theta$-temperature and the width of the transition region. For the study of the sodium chloride influence on the swelling, simulations with concentrations of 5 and 10~mol/l sodium chloride in water were performed with the three models. These simulations allow to identify suitable salt models.\\
Prior to these simulations, some preliminary studies were performed in which the different salt models were simulated in pure water. With these simulations, the thermophysical consistency of the models was tested. The used models for sodium chloride are listed in Table \ref{table5}.
\begin{table}[h!]
  \centering
  \caption{Lennard-Jones parameters ($\sigma$ and $\epsilon$) and point charge magnitude ($q_{el}$) of the salt models Chandrasekhar et al. \cite{Chandrasekhar_1984}, OPLS \cite{Chandrasekhar_1984,Aqvist_1990} and G53A6 \cite{Oostenbrink_2004_1}, where e is the elementary charge.}
  \label{table5}
  \begin{tabular}{ccccccc}
    \hline\noalign{\smallskip}
    model & \multicolumn{3}{c}{Na} & \multicolumn{3}{c}{Cl} \\
    \noalign{\smallskip}\hline \hline\noalign{\smallskip}
    & $\sigma$ / nm & $\epsilon$ / kJ$\cdot$mol$^{-1}$ & $q_{el}$ / e & $\sigma$ / nm & $\epsilon$ / kJ$\cdot$mol$^{-1}$ & $q_{el}$ / e \\
    \noalign{\smallskip}\hline\noalign{\smallskip}
    Chandrasekhar & 0.189744 & 6.724270 & 1.0 & 0.441724 & 0.492833 & -1.0 \\
    OPLS & 0.212645 & 0.076479 & 1.0 & 0.441724 & 0.492833 & -1.0 \\
    G53A6 & 0.257536 & 0.061749 & 1.0 & 0.444796 & 0.445708 & -1.0 \\
    \noalign{\smallskip}\hline
  \end{tabular}
\end{table}\\
For equilibration, the single PNIPAAm chain in water was simulated in the isobaric-isothermal ensemble over 1 to $5\cdot10^{7}$ time steps. The pressure was specified to be 1~bar and was controlled by the Berendsen barostat \cite{Berendsen_1984}, the time step was 1~fs for all simulations. Newton's equations of motion were numerically solved with the leap-frog integrator \cite{Hockney_1974}. For the long-range electrostatic interactions, particle mesh Ewald \cite{Essmann_1995} with a grid spacing of 0.12~nm and an interpolation order of four was used. A cutoff radius of $r_{c}=1.5$~nm was assumed for all interactions. After equilibration, $2$ to $4\cdot10^{7}$ production time steps were carried out with constant simulation parameters. Note that the production steps include the conformation transition as well as the simulation of the equilibrium.\\
In order to analyze the results, the radius of gyration $R_{g}$ was calculated
\begin{equation}
  R_{g}=\left(\frac{\Sigma_{i}||\textbf{r}_{i}||^{2}m_{i}}{\Sigma_{i}m_{i}}\right)^{1/2},
\end{equation}
which characterizes the degree of stretching of the single chain, where $m_{i}$ is the mass of site $i$ and $||\textbf{r}_{i}||$ is the norm of the vector from site $i$ to the center of mass of the single chain. The radius of gyration in equilibrium was calculated as the arithmetic mean over the last $5\cdot10^{6}$ time steps of the simulation together with its standard deviation. For the visualization of the results the open source program VMD \cite{Humphrey_1996} was used.

\subsubsection{Temperature}

\begin{figure}[b!]
  \centering
  \includegraphics[width=0.8\textwidth]{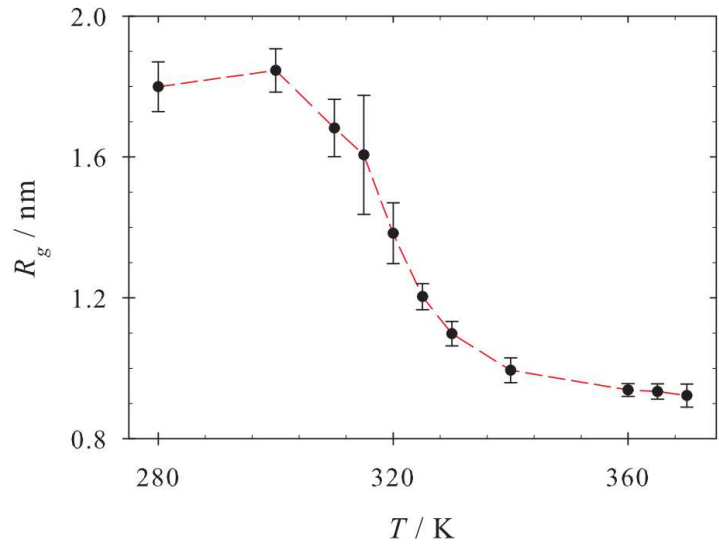}
  \caption{Radius of gyration $R_{g}$ of a PNIPAAm chain of 30 monomers in 14,482 water molecules in equilibrium over temperature $T$ for the force field combination OPLS + SPC/E. The error bars indicate the standard deviation.}
  \label{fig9}
\end{figure}

\begin{figure}[t!]
  \centering
  \includegraphics[width=0.8\textwidth]{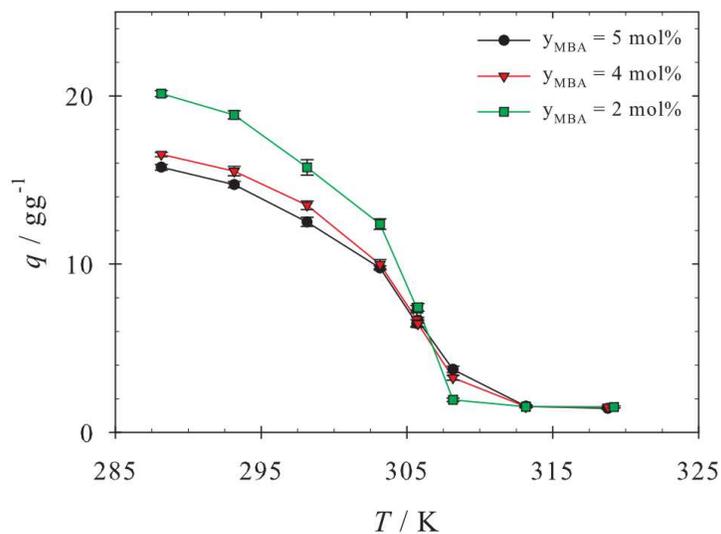}
  \caption{Degree of swelling of the PNIPAAm hydrogel with different amounts of cross-linker as a function of temperature. Symbols: experimental data, lines: guide for the eye. The error bars denote the standard deviation.}
  \label{exp}
\end{figure}

In a recent publication \cite{Walter_2010_2}, the results for the simulations and experiments of the swelling of PNIPAAm hydrogel over the temperature are described in detail. Here, the results are only shortly summarized.\\
For the successful force field combination OPLS + TIP4P, simulations at temperatures between 280 and 370~K were carried out in order to calculate the radius of gyration in equilibrium as a function of temperature and to use that information for determining the $\Theta$-temperature and the width of the transition region. The simulations were performed starting from the same stretched initial configuration. The radius of gyration in equilibrium was determined from a time average over the last 5~ns of the run.\\
Figure \ref{fig9} and Table \ref{table6} present the results for the radius of gyration as a function of the temperature for the force field combination OPLS + SPC/E. For temperatures below 300~K, the single chain is stretched, for temperatures above 340~K, it is collapsed. The $\Theta$-temperature is approximately 320~K. The width of the transition region is approximately 40~K.\\
\begin{table}[h!]
  \centering
  \caption{Radius of gyration $R_{g}$ of a PNIPAAm chain of 30 monomers in 14,482 water molecules in equilibrium over temperature $T$ for the force field combination OPLS + SPC/E. The numbers behind $\pm$ denote the standard deviation.}
  \label{table6}
  \begin{tabular}{cc}
    \hline\noalign{\smallskip}
    $T$ / K & $R_{g}$ / nm \\
    \noalign{\smallskip}\hline\noalign{\smallskip}
    280 & 1.80 $\pm$ 0.07 \\
    300 & 1.85 $\pm$ 0.06 \\
    310 & 1.68 $\pm$ 0.08 \\
    315 & 1.61 $\pm$ 0.17 \\
    320 & 1.38 $\pm$ 0.09 \\
    325 & 1.20 $\pm$ 0.04 \\
    330 & 1.10 $\pm$ 0.03 \\
    340 & 0.99 $\pm$ 0.04 \\
    360 & 0.94 $\pm$ 0.02 \\
    365 & 0.93 $\pm$ 0.02 \\
    370 & 0.92 $\pm$ 0.03 \\
    \noalign{\smallskip}\hline
  \end{tabular}
\end{table}\\
With the force field combination OPLS + SPC/E, qualitatively correct results were achieved. The main difference between the experimental data (cf. Figure \ref{exp}) and the prediction by molecular simulation is that the $\Theta$-temperature is about 15~K higher and the width of the transition region is overestimated by the molecular simulation. In summary, this is an unexpectedly favorable agreement between the prediction by molecular simulation and the experimental data, especially when considering that the force fields were not trained on such type of data and no adjustments were made.
\\

\subsubsection{Sodium chloride concentration}

\begin{figure}[h!]
  \centering
  \includegraphics[width=0.8\textwidth]{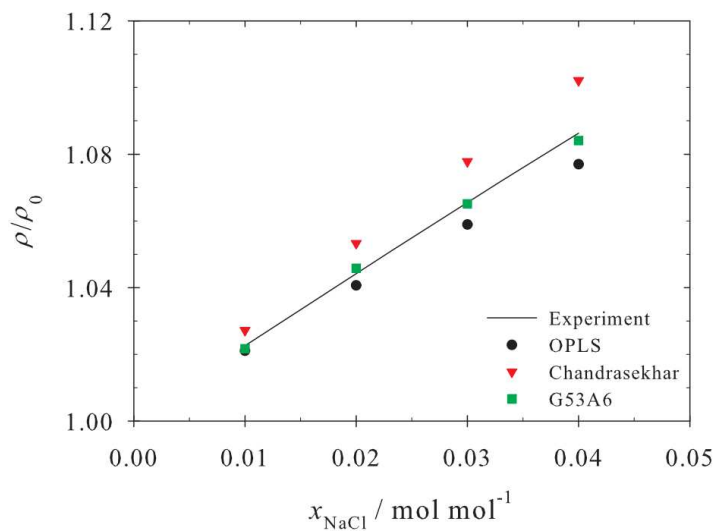}
  \caption{Reduced density $\rho/\rho_{0}$ of an aqueous sodium chloride solution as function of the salt concentration, given in mole fraction $x_{NaCl}$ for different sodium chloride models at 398~K in SPC/E water. The black line denotes experimental data from Zhang and Han \cite{Zhang_1996}}
  \label{salt}
\end{figure}

\begin{figure}[b!]
  \centering
  \includegraphics[width=0.8\textwidth]{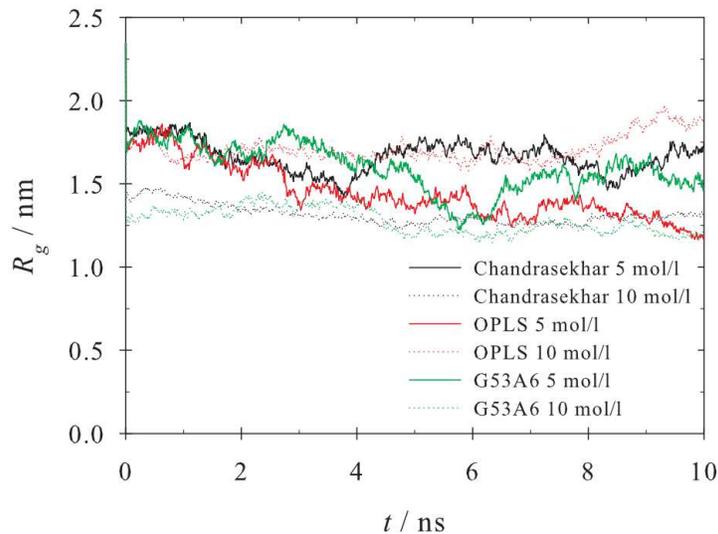}
  \caption{Radius of gyration $R_{g}$ of a PNIPAAm chain of 30 monomers in 14,482 water molecules over simulation time $t$ for the force field combination OPLS + SPC/E with different sodium chloride models at 300~K.}
  \label{rg_salt}
\end{figure}

\begin{figure}[h!]
  \centering
  \includegraphics[width=0.8\textwidth]{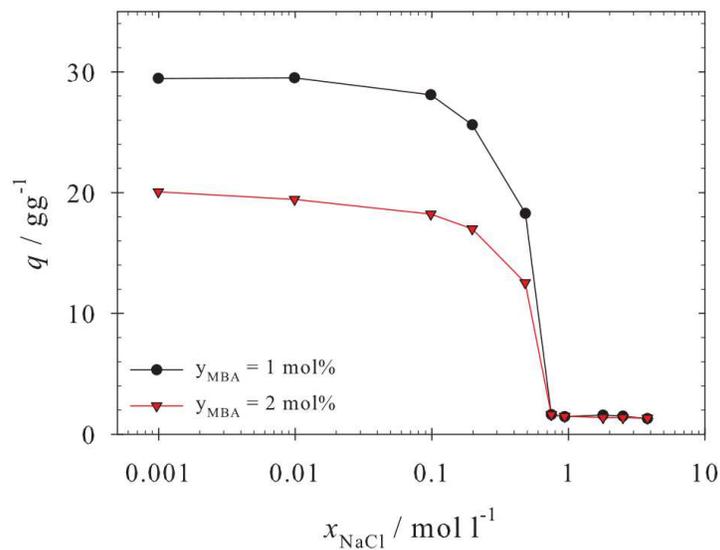}
  \caption{Degree of swelling of the PNIPAAm hydrogel with different amounts of cross-linker as a function of sodium chloride concentration. Symbols: experimental data, lines: guide for the eye.}
  \label{q_salt}
\end{figure}

In the preliminary study, the three models for sodium chloride were simulated in SPC/E water at 298~K for different salt concentration. In these simulations the density of the electrolyte solution was measured. For comparison with experimental data \cite{Zhang_1996}, the density was reduced by the density of pure water ad plotted over the mole fraction of sodium chloride. The results can be seen in Figure \ref{salt}. It is obvious that the G53A6 model describes the experiment best. The results can be interpreted by looking on the water models the sodium chloride models were developed with: Chandrasekhar was developed with TIP4P \cite{Jorgensen_1983}, OPLS with TIP4P/SPC (two different models were used) and G53A6 with SPC \cite{Berendsen_1981}.
\\
In clarify, which force field is able to predict the swelling of PNIPAAm in aqueous solutions of sodium chloride, a single PNIPAAm chain of 30 monomers was simulated in water with 5 and 10~mol/l of sodium chloride. The results as radius of gyration over the simulated time are plotted in Figure \ref{rg_salt}. The simulations were started throughout with a stretched initial configuration of the single chain. The salt models G53A6 and Chandrasekhar both yield a collapsed conformation in equilibrium for 10~mol/l and a conformation in between the stretched and the collapsed one for 5~mol/l. The model OPLS yields the opposite result. In experiments, the PNIPAAm hydrogel is swollen in pure water and collapsed at high of sodium chloride concentrations (cf. Figure \ref{q_salt}).
\\

\subsubsection{Computational demand}

\begin{figure}[t!]
  \centering
  \includegraphics[width=0.8\textwidth]{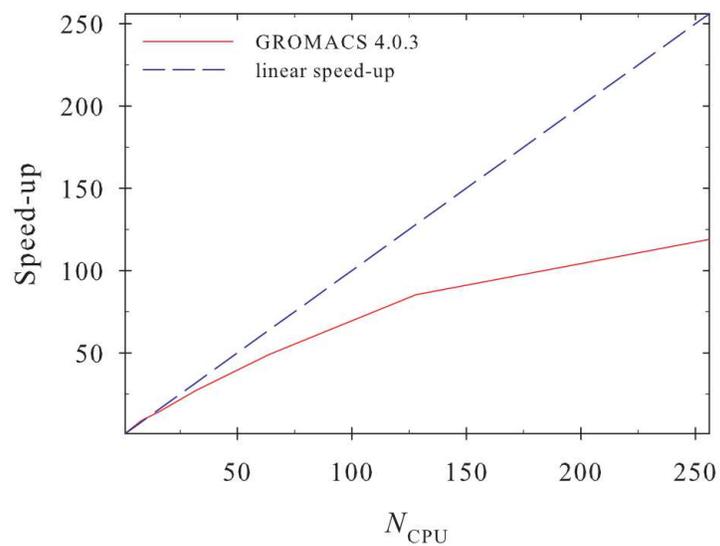}
  \caption{Weak scaling of the massively parallel program GROMACS 4.0.3 on HP~XC4000, simulation time reduced with the simulation time on one processor over number of processors. A PNIPAAm chain with 30 monomers in 14~482 water molecules was simulated.}
  \label{scaling}
\end{figure}

All presented simulations in Section \ref{hydrogel} were carried out with the MPI based molecular simulation program GROMACS. The parallelization of the molecular dynamics part of GROMACS is based on the eighth shell domain decomposition method \cite{Hess_2008}. With GROMACS, typical simulation runs to determine the radius of gyration in equilibrium employ 64--128 CPUs running for 1--3 days. For these simulations very large systems must be considered comprising typically about 58~800 interaction sites. Table \ref{scaling} demonstrates the good scaling of the program on the HP~XC4000 cluster at the Rechenzentrum of the Universit\"at Karlsruhe (TH). For these simulations a maximum memory of 284~MB and a maximum virtual memory of 739~MB was used.\\

\subsection{Nucleation of fluids}

The key properties of nucleation processes are the nucleation rate $J$ that indicates how many clusters of the emerging phase appear in a given volume per time and the height $\Delta\Omega*$ of the free energy barrier that must be overcome to form stable nuclei. The most widespread approach for calculating these quantities is the classical nucleation theory \cite{Feder}, which has significant shortcomings, e.g., it overestimates $\Delta\Omega*$ for homogeneous vapor to liquid nucleation \cite{Talanquer}. A more accurate theory of homogeneous nucleation, which is sought after, would also increase the reliability for more complex applications such as heterogeneous and ion-induced nucleation in the earth’s atmosphere.\\
An important problem is that the basic assumptions underlying the classical approach do not apply to nanoscopic nuclei \cite{Merikanto}. Properties of such nuclei are hard to investigate experimentally, but are well accessible by molecular simulation. For instance, equilibria \cite{Vrabec} and vaporization processes \cite{Holyst} of single liquid droplets can be simulated to obtain the surface tension as well as heat and mass transfer properties of strongly curved interfaces. Similarly, very fast nucleation processes that occur in the immediate vicinity of the spinodal are experimentally inaccessible, whereas they can be studied by Monte Carlo \cite{Neimark} and molecular dynamics \cite{Vrabec_2} simulation of systemswith a large number of particles. Hence, molecular simulation is crucial for the further development of nucleation theory.\\
In the recent investigations \cite{Horsch_1,Horsch_2}, grand canonical molecular dynamics with McDonald’s daemon is discussed and applied for sampling both nucleation kinetics and steady-state properties of a supersaturated vapour. Here, a series of simulations is conducted for the truncated and shifted Lennard-Jones fluid which accurately describes the fluid phase coexistence of noble gases and methane.

\section{Conclusion}

In Section \ref{dev} it was shown, how molecular models can be developed. Two new models that carbon dioxide and cylcohexanol were introduced. Both models are able to predict many different properties they were not adjusted to quantitatively correct.\\
For the model development, also a new procedure via reduced units was introduced and succesfully used in the development of the model for cyclohexanol.\\
\\
In Section \ref{ver}, the verification and usage of force fields and models for different applications was discussed.\\
The temperature dependence of swelling of PNIPAAm hydrogels in pure water was studied both experimentally and by molecular simulation.\\
It was found that it is possible to study the conformation transition of PNIPAAm hydrogel by molecular simulation of the single polymer chain. The force field combinations G53A6 + TIP4P and OPLS + SPC/E yield the stretched and the collapsed conformations as well as the conformation transition. Four other studied force field combinations yield only the collapsed conformation. With the force field combination OPLS + SPC/E, it is possible to qualitatively predict the swelling of the hydrogel as a function of temperature. A $\Theta$-temperature for the PNIPAAm single chain in pure water of approximately 320~K and a width of the conformation region of approximately 40~K was predicted.\\
All three tested models for sodium chloride, namely Chandrasekhar, OPLS and G53A6, were able to describe the density of aqueous electrolyte solutions qualitatively. With the model G53A6 it was also possible to describe the density quantitatively. Different conformations of the PNIPAAm hydrogel as a function of sodium chloride concentration in water could be described with all three models for the salt. But the OPLS force field describes a stretching of the chain for high sodium chloride concentration of 10~mol/l. This is the opposite of what the experiment shows.\\
Molecular simulation is a useful tool for predicting conformation transitions of real hydrogels. Accurate quantitative results can be expected, if suitable force fields are available.\\
It was also shown that it is possible to simulate the nucleation of real fluids and determine the nucleation rate from these simulations. Grand canonical molecular dynamics with McDonald’s daemon was established as a method for steady-state simulation of nucleating vapours at high supersaturations. A series of simulations was conducted for the truncated and shifted Lennar-Jones fluid. The classical nucleation theory was found to underpredict the nucleation rate below the triple point.

%


\end{document}